\documentclass{article}
\usepackage{emulateapj,epsfig}

\begin{document}

\newcommand{\totder}[2]{\frac{d{#1}}{d{#2}}}
\newcommand{\parder}[2]{\frac{\partial{#1}}{\partial{#2}}}

\newcommand{\gsim}{\mbox{\hspace{.2em}\raisebox{.5ex}{$>$}\hspace{-.8em}\raisebox{-.5ex}{$\sim$}\hspace{.2em}}}
\newcommand{\lsim}{\mbox{\hspace{.2em}\raisebox{.5ex}{$<$}\hspace{-.8em}\raisebox{-.5ex}{$\sim$}\hspace{.2em}}}
\newcommand{\beq}{\begin{equation}}     \newcommand{\eeq}{\end{equation}}
\newcommand{\bey}{\begin{eqnarray}}     \newcommand{\eey}{\end{eqnarray}}
\newcommand{\etal}{{\em et al.\/}}
\newcommand{\ie}{{\em i.e.\/}}  \newcommand{\eg}{{\em e.g.\/}}
\newcommand{\lt}{\left} \newcommand{\rt}{\right}
\newcommand{\ssst}{\scriptscriptstyle}
\newcommand{\E}[1]{\times 10^{#1}}

\newcommand{\FIR}{\,{\rm FIR}}
\newcommand{\B}{\,{\rm B}}
\newcommand{\s}{\,{\rm s}}      \newcommand{\ps}{\,{\rm s}^{-1}}
\newcommand{\yr}{\,{\rm yr}}    \newcommand{\pHz}{\,{\rm Hz}^{-1}}
\newcommand{\cm}{\,{\rm cm}}    \newcommand{\km}{\,{\rm km}}
\newcommand{\kmps}{\km\s^{-1}}
\newcommand{\parsec}{\,{\rm pc}}\newcommand{\Mpc}{\,{\rm Mpc}}
\newcommand{\ergs}{\,{\rm ergs}}        \newcommand{\K}{\,{\rm K}}
\newcommand{\eV}{\,{\rm eV}}    
\newcommand{\nufnu}{$\nu f_{\nu}$}

\newcommand{\za}{z_{1}}         \newcommand{\zb}{z_{2}}
\newcommand{\zo}{z_{0}}         \newcommand{\tha}{\theta_{app}}
\newcommand{\va}{v_{1}}         \newcommand{\vb}{v_{2}}
\newcommand{\Ys}{Y_{\ast}}
\newcommand{\cs}{\xi_{\ast}}    \newcommand{\ts}{\theta_{\ast}}
\newcommand{\Dy}{\Delta_{Y}}    \newcommand{\Dv}{\Delta_{v}}
\newcommand{\ro}{r_{\ssst 0}}
\newcommand{\Ha}{\mbox{H$\alpha$}}

\title{A Possible Evolutionary Connection between AGN and Starburst in LINERs}

\author{S.J. Lei\altaffilmark{1,2}, J.H. Huang\altaffilmark{1,2}, W. Zheng\altaffilmark{3}, L. Ji\altaffilmark{1,2}, Q.S. Gu\altaffilmark{1,2}}
\altaffiltext{1}{Department of Astronomy, Nanjing University, Nanjing, China, lsj,jhh,qsgu@nju.edu.cn}
\altaffiltext{2}{United Lab for Optical Astronomy, The Chinese Academy of Sciences, China}
\altaffiltext{3}{Department of Physics and Astronomy, The Johns Hopkins University, Baltimore, MD 21218-2686, U.S.A., zheng@pha.jhu.edu}

\begin{abstract}
Our analysis on the two magnitude-limited samples of LINERs suggests  
a correlation between $L_{\FIR}/L_{\B}$ or f(25\micron)/f(60\micron), 
and Hubble-type index at $>$99.99\% 
significance level.  As $L_{\FIR}/L_{\B}$ and f(25\micron)/f(60\micron)~
are considered as the indicators of star-formation activity and AGN activity, 
respectively, our result suggests that LINERs with higher AGN activity may 
have a lower star-formation contribution. 
The ones with highest AGN activity and lowest star-formation contributions are
ellipticals. All well-studied LINER 1s belong to this group. On the other
hand, LINERs with higher star-formation activity present lower AGN
contributions. We find all well-studied LINER 2s in this parameter space. 
Most of LINERs having inner ring structures belong to this group.
Statistics with other indicators of star-formation or AGN activity
(nulear mass-to-light ratio at H band, and the ratio of X-ray-to-UV power)
provide further evidence for such a trend. We have seen that along with 
the evolution of galaxies from late-type spirals to early-type ones, and up 
to ellipticals, the intensity of AGN activity increases with
decreasing star-formation contributions, The above analyses 
may suggest a possible connection between the host galaxies and nuclear 
activities, and it might also indicate a possible evolutionary connection
between AGN and starburst in LINERs.

\end{abstract}

\section{Introduction}
 It was the study of  Simkin, Su, and Schwarz (1980) that
 revealed a more frequent occurrence of outer and inner rings in 
the host galaxies of Seyferts than those of non-Seyfert galaxies. This 
pioneering investigation has been confirmed recently (Hunt \& Malkan 1999) 
of the possible connection between the host galaxies and nuclear activities.

The galaxy morphology can be modified by environmental effects, such as
mergers or interactions. It could be formed, on the other hand, as an
evolutionary sequence (Martinet 1995), without galaxy
interactions.

The role of minor mergers in the formation of AGN has been tested
 by Corbin (2000) . A very  intriguing  result obtained from this
 study is that the nuclear spectral type of galaxies is strongly
 dependent on their Hubble type instead of environmental effects,
 indicating the existence of  a connection between the host galaxy and
 nuclear activity.

 When Condon et al. suggested (1982), and Rush et al. (1996) conducted,
 to distinguish
 starbursts from AGN by   using  of FIR-radio correlation, the possible
 connection between the host galaxy and nuclear activity  is
 involved. The nuclear spectral types of sources can be designated by
 their global properties. The FIR-radio correlation has proven a useful tool
 successfully classifying the LINERs' type, LINER 1s and LINER 2s, in
 recent investigation (Ji et al.2000). In their study on a small sample
 of LINERs,  a  higher frequency of inner rings in LINERs found by Hunt \&
 Malkan (1999) turned out to be related to LINER 2s.

It is worth exploring the astrophysical implication of this result,
especially for understanding a possible connection between the host
galaxy and nuclear activity, or even further a possible connection
between starbursts and AGN in terms of the evolution  of galaxies.

\section{LINERs' sample}

To faciliatate the study of active nuclei and their hosts, it is desirable to
utilize a LINER sample on the basis of host galaxy flux (Krolik 1999). 
One of the best sample in the literature is the spectroscopic survey by 
Ho, Filippenko and Sargent (1997, HFS hereafter), from which we have a 
magnitude-limited LINER sample, with a statistically meaningful size of 94.

A second, and deeper, magnitude-limited LINER sample has been constructed
from a catalog of LINER compiled by Carrillo et al. (1999, CMD hereafter), 
which is  based on criteria of m$_{\rm B} \leq$ 14.5 and $\delta \geq 
0^{\circ}$. The CMD sample contains 223 sources.

%

\section{Statistical Results}
%
%
%
\subsection{Star Formation and AGN Activity}

In order to test a possible connection between the host galaxies and
nuclear activities, we need
to study the star-formation activity among LINER 
samples, along with their AGN activity. As an indicator of
star-formation activity, the luminosity
ratio of $L_{\FIR}/L_{\B}$ has been widely used (Keel 1993, Huang et al. 1996,
Hunt \& Malkan 1999).
On the other hand, the mid-infrared flux ratio f(25\micron)/f(60\micron)
 has been known a good indicator of AGN activity (Hunt \& Malkan 1999).

Following Hunt \& Malkan (1999), we make plots of f(25\micron)/f(60\micron)
vs Hubble-type index for the HFS sample and the CMD sample in Fig 1a and 1b
respectively. 
Significant correlations between them, $>$99.99\% for the 
HFS sample and $>$99.99\% for the CMD sample, imply a possible 
relation of nuclear activity
to the evolutionary status of galaxy, similar to the results obtained by
Corbin (2000) and Hunt \& Malkan (1999) for AGN. 

The different distributions shown in Fig 1a and 1b lie mostly in the lack of
LINERs in late-types for the HFS sample, as compared with those for the CMD
sample. It is something related to the selection effect in the HFS sample,
which makes it harder to find AGN in late-type galaxies. 
Though the CMD sample is not complete for sources with 
12.5 $< m_{\rm B} \leq$ 14.5, the situation is much improved. 
This selection effect may cause some problems on the study of 
evolutionary status of AGN activity along the Hubble sequence. However, it is 
not crucial for our investigation as the statistics given above indicated.  

In Fig 2 we have shown the star-formation indicator 
of $L_{\FIR}/L_{\B}$ vs Hubble-type index for the CMD sample only. The 
correlation is significant at the level of $>$99.99\%. What the trend in Fig 2
shows is certainly consistent with the fact that 
Hubble-type sequence is also a sequence of star formation rate (Kennicutt 1992).
The most active star formation occurs in late-type spirals. In fact, 
the majority of LINERs with inner rings are located in these types, see Fig 1
and 2.


\subsection{Further Test}
There are other significant indicators for star-formation or AGN activity, 
such as the nuclear mass-to-light ratio at H band, $M/L_{\rm H}$, and
the ratio of X-ray-to-UV power, \nufnu(2-10kev)/\nufnu(1300\AA) (Oliva et al. 
1999; Maoz et al. 1998).

A diagram of $M/L_{\rm H}$  vs f(25\micron)/f(60\micron) for Seyferts 
is shown in Fig 3a with the $M/L_{\rm H}$ data obtained by Oliva et al. (1999),
which again demonstrates the f(25\micron)/f(60\micron) ratio as a good 
indicator of AGN activity. The sources in the lower-left part of the 
diagram are Seyfert 2 with circum-nuclear starbursts, while those 
in the upper-right part are Seyfert 1 without circum-nuclear starbursts.
The trend shown in Fig 3a might provide evidence for the evolutionary 
hypothesis for Seyferts (Hunt \& Malkan 1999).

Following this approach, we made a similar plot for LINERs in Fig 3b, 
with the $M/L_{\rm H}$ data obtained by Devereux et al. (1987). The correlation
between $M/L_{\rm H}$  and f(25\micron)/f(60\micron) is significant at 97\%
level. Two sources shown with symbol of star, NGC1052 and NGC4486, are those
with the $M/L_{\rm H}$ data obtained by Oliva et al. (1999). The systematic 
difference between the two observations is obvious, as it can be seen from 
different positions of the same object, NGC1052. 

Further evidence for this evolutionary trend might be provided using
another AGN activity, i.e. the ratio of X-ray-to-UV power. Due to
the limited LINERs with both X-ray detections and the UV observations, we 
 use the Einstein data (Carrillo et al. 1999) for the X-ray fluxes instead of
the hard X-ray fluxes used by Maoz et al. (1998). The UV fluxes are retrieved
from the IUE archive data.\footnote{Based on INES data from the IUE satellite}
The correlation between 
$L_{\FIR}/L_{\B}$ and \nufnu(0.2-4.0kev)/\nufnu(1300\AA) is shown in Fig 4, 
significant at 99\% level. 

\section{Discussions}
A key point in AGN unification hypotheses is to claim that Seyfert 1s and 2s
are intrinsically similar, and the different types of nuclear activity are
caused by different view angles.  Increasing evidence
(Malkan et al. 1998; Hunt et al. 1999) suggests that Seyfert nuclei
may be same objects seen at different evolutionary sequence. Seyfert 2s
tend to reside in later morphological types than Seyfert 1s. Starbursts
and massive stars play an important energetic role in a significant
fraction of Seyfert 2s (Heckman 1999), while few Seyfert 1s have such
circum-nuclear starbursts (Gonzalez Delgado et al. 1997; Hunt et al. 1997).

The data points in the lower right part of Fig. 4
of the diagram have higher ratio of \nufnu(0.2-4.0kev)/\nufnu(1300\AA)
and lower ratio of $L_{\FIR}/L_{\B}$, in other words, have higher 
AGN activity and lower star-formation contributions. It is the region 
the well studied LINERs with broad \Ha~ emission or with active AGN 
are located at, e.g. NGC3998, NGC4486, NGC4594(Larkin et al. 1998;
Ho 1998; Nicholson et al. 1998;
Maoz et al. 1998). We have found few LINERs with inner rings in this region.
Sources with the highest AGN activity and the lowest star-formation 
contributions are absolutely ellipticals. It has been known that most of
X-rays detected in ellipticals come from the extended halo where elliptical
galaxies reside. Removing these ellipticals from the sample used
in Fig 4, the correlation between $L_{\FIR}/L_{\B}$ and
\nufnu(0.2-4.0kev)/\nufnu(1300\AA) becomes significant at 90\% level.

On the contrary, the sources shown in the upper left part of the figure
have lower ratio of \nufnu(0.2-4.0kev)/\nufnu(1300\AA) and higher ratio of
$L_{\FIR}/L_{\B}$, i.e. lower AGN contribution and higher star-formation
activity. It is the region some well studied LINERs supported by
massive stars are found, e.g. NGC4569, NGC4736, NGC4826, NGC5194 (Maoz et al.
1998; Larkin et al. 1998; Alonso-Herrero et al. 1999; Barth \& Shields 2000). 

The above argument holds for Fig. 3b, too. The AGN-supported LINERs,
NGC1052, NGC4579, NGC4486, are distributed in the region with 
higher ratio of f(25\micron)/f(60\micron) (strong AGN activity)
and higher nuclear $M/L_{\rm H}$ (small starburst contribution).
While NGC7217, a well studied starburst(SB)-supported LINER, is in the 
region having higher star-formation activity and lower AGN contribution.

If adopting the Hubble sequence as an evolutionary sequence (Pfenniger, Combes 
\& Martinet 1994, Martinet 1995), the correlations in Fig 1 and 
2 show that the AGN and the star-formation activity are anticorrelated
to the evolutionary status of galaxies. Along with the evolution from
late-type spirals to early-type ones, and up to ellipticals, the intensity 
of AGN activity in LINERs increases with decrease
of star-formation contributions, consistent with what the statistics 
performed for Fig 3 and 4 implied. Due to the large sample size used for
Fig 1 and 2, more well studied LINERs, AGN- or SB-supported, can be found
in their corresponding regions, e.g. SB-supported LINERs NGC404, NGC3504,
NGC5055, NGC7743 (Maoz et al. 1998; Alonso-Herrero et al. 1999; Larkin et al
1998), and AGN-supported ones NGC2639, NGC3718, NGC4203, NGC4278, NGC6500
(Alonso-Herrero et al. 1999; Ho 1998; Iyomoto et al. 1998; Shields et al. 2000;
Falcke et al. 1999; Maoz et al. 1998).

Probably, this evolutionary trend might indicate processes in which 
massive black holes in LINERs are competing with the starbursts for 
the inflowing gas, a suggestion proposed by de Carvalho \& Coziol (1999).

The LINERs residing in late-type spirals, with strong star-formation 
and lower AGN activity, might be at an early evolutionary stage. They are 
LINER 2s. The growing bulges of galaxies are bound to be followed by the
evolution of galaxies from late-type to early-type spirals.
The formation of a massive black hole may turn out to be a
natural evolution of the massive bulges of galaxies (Norman \& Scoville
1988). Ellipticals, S0 galaxies and very early-type spirals
containing LINERs with broad \Ha~ emission or active AGN have
massive bulges for the formation of massive or supermassive black holes
(Wandel 1999; Margorrian et al. 1998), 
being active enough to suppress the circum-nuclear starbursts in LINERs.
These LINERs might be at late evolutionary stages of LINERs, they are LINER 1s.

In conclusion, the above analyses
may suggest a possible connection between the host galaxies and nuclear
activities, and it might also indicate a possible evolutionary connection
between AGN and starburst in LINERs.


\acknowledgements{The authors are very grateful to the anonymous referee 
for his critical comments, a significant revision was made according to his 
suggestions. We are also thankful to Dr. Tinggui Wang for help with the 
IUE data retrieval.
This work is supported by a grant from the NSF of China,
and a grant from the Basic Research Project of the State Scientific Commission 
of China.}

\clearpage

\begin{figure*}
\figurenum{1}
\plotone{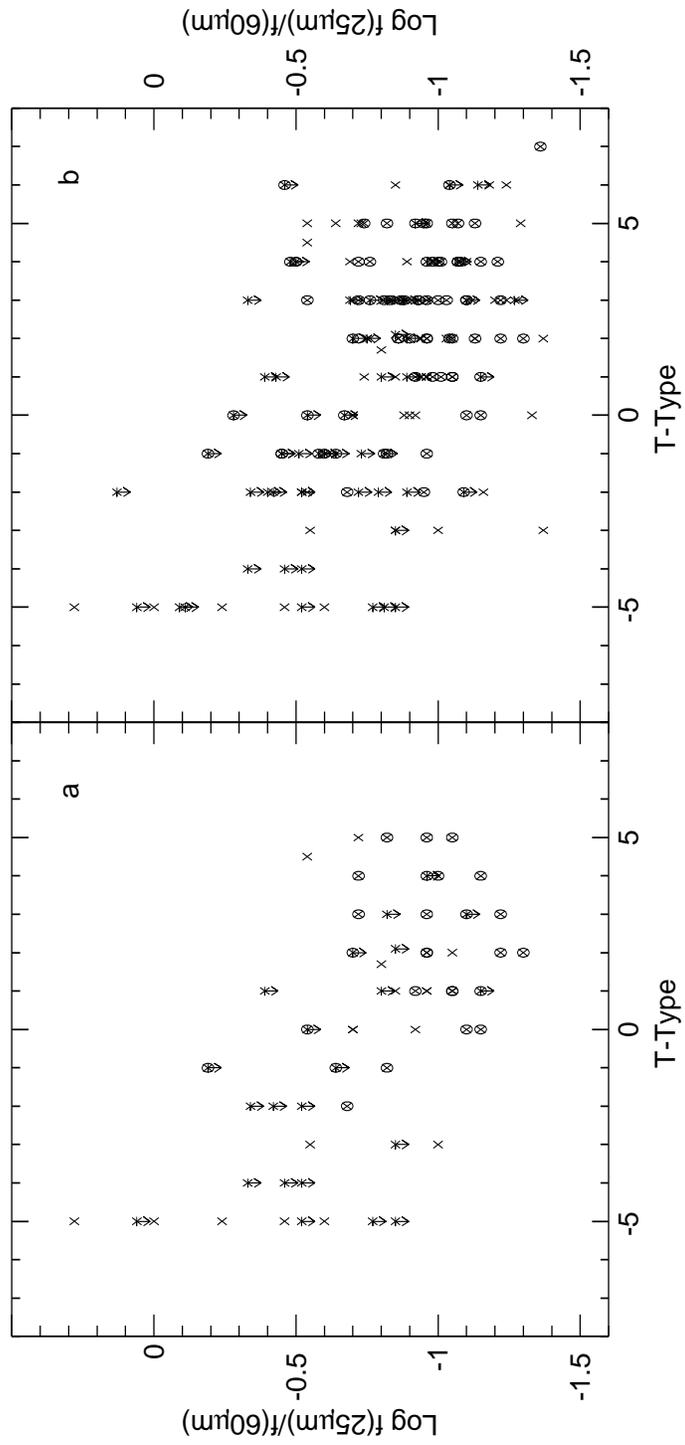}
\caption{Distributions of f25\micron/f60\micron, indicator of AGN activity,
along the Hubble-type index T for a) the HFS sample; b) the CMD sample.
Symbols enclosed with open circles are for LINERs with inner rings. Censored
data are indicated by arrows attached to the data points.}
\end{figure*}

\begin{figure*}
\figurenum{2}
\plotone{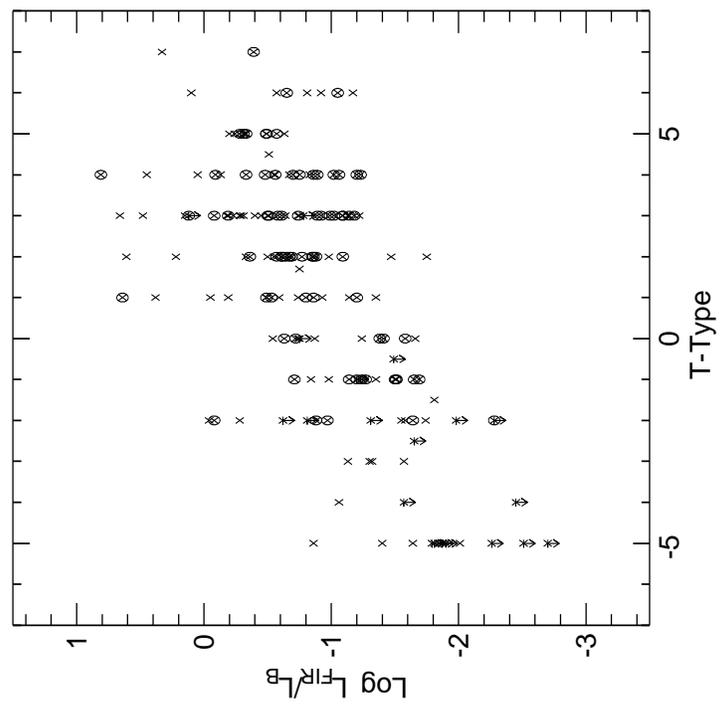}
\caption{Diagram of $L_{\FIR}/L_{\B}$ vs Hubble-type index T for CMD sample
with same denotations as those in Fig 1.}
\end{figure*}

\begin{figure*}
\figurenum{3}
\plotone{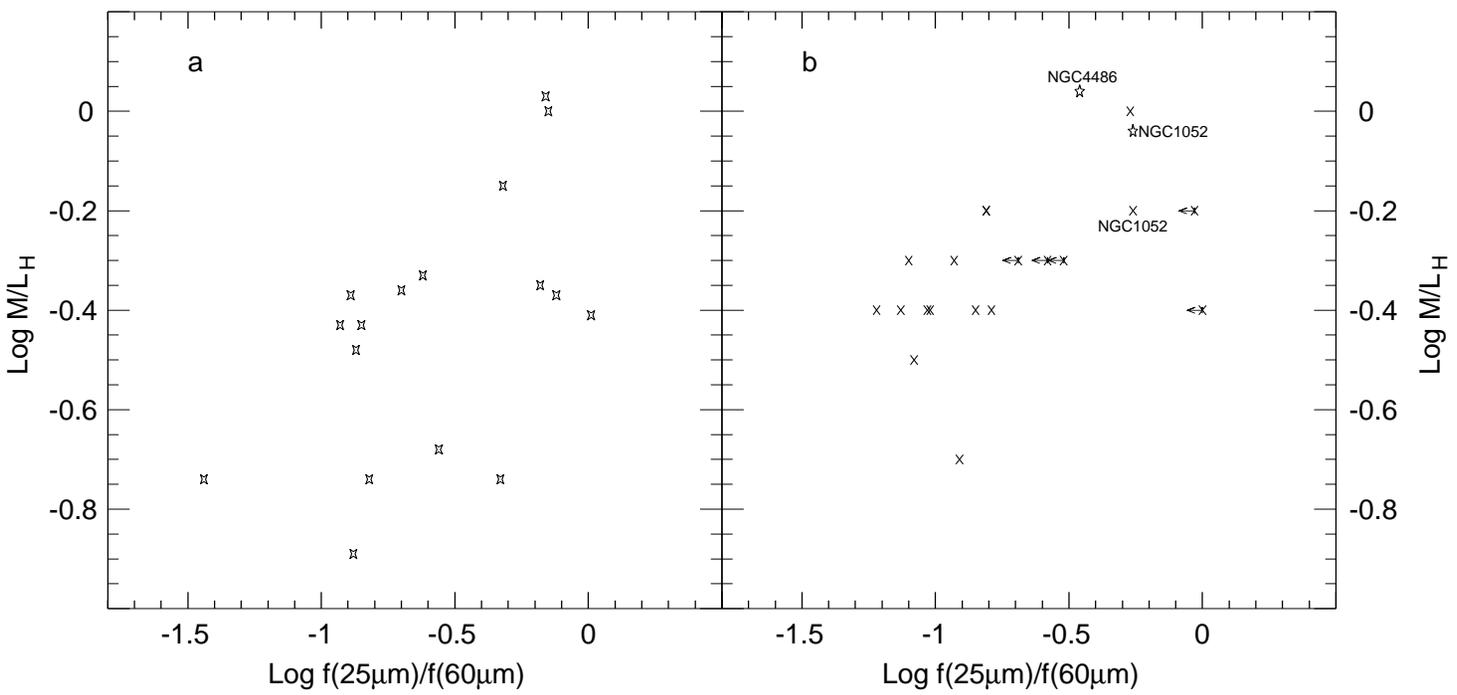}
\caption{Plots of nuclear mass-to-light ratio at H band, $M/L_{\rm H}$ vs
f25\micron/f60\micron for a) Seyferts; b) LINERs. Censored data are indicated
by arrows attached to the data points.}
\end{figure*}

\begin{figure*}
\figurenum{4}
\plotone{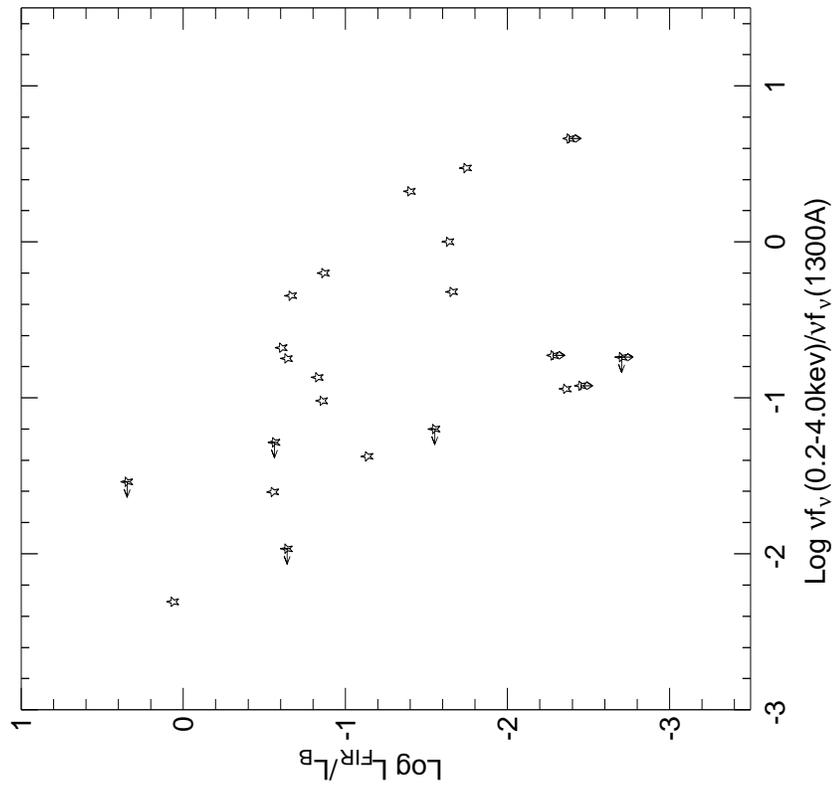}
\caption{Diagram of $L_{\FIR}/L_{\B}$ vs \nufnu(0.2-4.0kev)/\nufnu(1300\AA)
for LINERs. Censored data are indicated by arrows attached
to the data points.}
\end{figure*}

\end{document}